\documentstyle[aps,12pt]{revtex}
\begin{document}
\title{Shape Coexistence in the Relativistic Hartree-Bogoliubov approach}
\author{T. Nik\v si\' c$^{1,2}$, D. Vretenar$^{1,2}$, P. Ring$^{2}$, and
G.A. Lalazissis$^{2,3}$}
\address{$^{1}$ Physics Department, Faculty of Science, University of\\
Zagreb, 10 000 Zagreb, Croatia\\
$^{2}$ Physik-Department der Technischen Universit\"at M\"unchen,\\
D-85748 Garching, Germany\\
$^{3}$Department of Theoretical Physics, Aristotle University of\\
Thessaloniki,\\
Thessaloniki GR-54006, Greece\\
}
\maketitle
\vspace{1cm}
\begin{abstract}
The phenomenon of 
shape coexistence is studied in the Relativistic Hartree-Bogoliubov
framework. Standard relativistic mean-field effective interactions do not
reproduce the ground state properties of neutron-deficient Pt-Hg-Pb
isotopes. It is shown that, in order to consistently describe binding
energies, radii and ground state deformations of these nuclei, effective
interactions have to be constructed which take into account the sizes of
spherical shell gaps. 
%#####################################################
\end{abstract}

\bigskip

\bigskip \bigskip

%#####################################################
\bigskip \bigskip

%#####################################################
\vspace{1 cm} {PACS:} {21.60.Jz, 21.10.Dr, 27.70.+q, 27.80.+w}\newline
\vspace{1 cm}\newline
\newpage \baselineskip = 24pt

%=========================================================================
%  Section 1
%

\section{Introduction}

%=========================================================================
Nuclear structure models based on the relativistic mean-field approximation
have been successfully employed in the description of ground-state
properties of nuclei all over the periodic table. By adjusting a minimal set
of model parameters, meson masses and coupling constants, to binding
energies and radii of few spherical closed shell nuclei, it has been
possible to perform detailed structure calculation of a large number of
spherical and deformed nuclei~\cite{Rin.96}. More recently, the relativistic
mean-field framework has also been used in studies of nuclear structure
phenomena in exotic nuclei far from the valley of $\beta$- stability and of
the physics of the drip lines. In particular, the Relativistic
Hartree-Bogoliubov model encloses a unified description of mean-field and
pairing correlations, and we have used this model to calculate ground-state
properties of exotic nuclei with extreme isospin values: the halo phenomenon
in light nuclei~\cite{PVL.97}, properties of light
neutron-rich nuclei~\cite{LVP.98}, the reduction of the effective single-nucleon spin-orbit potential
in nuclei close to the drip-lines~\cite{LVR.97}, properties of neutron-rich
Ni and Sn isotopes~\cite{LVR.98}, the location of the
proton drip-line from $Z=31$ to $Z=73$ and the phenomenon
of ground-state proton radioactivity ~\cite{VLR.99,LVR.99,LVR.01b}.

On the other hand, it is still an open problem how far from stability can
one extrapolate the predictions of standard relativistic mean-field
effective interactions. The question is how well can effective interactions,
which are adjusted to global properties of spherical closed shell nuclei,
describe the structure of nuclei far from stability, or predict new
phenomena in nuclei with extreme isospin values. It is well known, for
example, that various non-relativistic and relativistic mean-field models
differ significantly in the prediction of the exact location of the neutron
drip-line. This is, of course, related to different isovector
properties of various effective interactions. In general, on cannot expect
standard mean-field models to accurately describe the properties which
crucially depend on the proton and neutron single-particle spectra. An
important example is the suppression of shell effects and the related
phenomenon of deformation and shape coexistence.

A number of theoretical analyses of shell quenching and shape coexistence
phenomena have been performed in the relativistic mean-field framework. In
Ref.~\cite{Lal.99} we have studied the dissolution of the spherical N=28
shell gap in neutron-rich nuclei. By performing constrained Relativistic
Hartree-Bogoliubov calculations, we have shown how the strong reduction of
the gap between the neutron $1f_{7/2}$ orbital and the $2p_{3/2}$, $2p_{1/2}$
levels results in deformed ground states and shape coexistence in
neutron-rich Si, S and Ar isotopes, in excellent agreement with experimental
data. On the proton-rich side, the relativistic mean-field model has been
employed in studies of shape coexistence in neutron-deficient Pt, Hg and Pb
isotopes~\cite{Yos.94,Pat.94,Tak.96,Yos.97}. However, as has been pointed
out in Ref.~\cite{Hey.96}, the results of these studies were at variance
with experimental data, especially their predictions of deformed ground
states in Pb isotopes, and of prolate and superdeformed ground states in Hg
isotopes. In a very recent analysis of the $Z=82$ shell closure in
neutron-deficient Pb isotopes~\cite{Ben.01}, relativistic mean-field
calculations again predicted deformed ground-states in some Pb isotopes, in
contradiction with experiment.

In this work we employ the Relativistic Hartree-Bogoliubov model in the
analysis of shape coexistence phenomena in neutron-deficient Hg and Pb
isotopes. Although standard relativistic mean-field forces do not reproduce
the experimental data on ground state properties, it is indeed possible to
construct effective interactions which consistently describe binding
energies, radii and quadrupole deformations of neutron-deficient nuclei in
this mass region. In addition to bulk properties, also the sizes of
spherical gaps in the single-nucleon spectra have to be taken into account
when adjusting the parameters of such an effective interaction. The
spherical magic and semi-magic gaps determine the relative excitation
energies of coexisting minima based on different intruder configurations.

Section II contains an outline of the Relativistic Hartree-Bogoliubov model.
The problem of the relativistic effective interaction and shape coexistence
in neutron deficient Hg and Pb isotopes is discussed in Section III. The
results are summarized in Section IV. 
%=========================================================================
%  Section 2
%

\section{Outline of the Relativistic Hartree-Bogoliubov Model}

%=========================================================================
In the relativistic framework a nucleus is described as a collection of
nucleons that interact by the exchange of effective mesons. The lowest order
of the quantum field theory is the {\it mean-field} approximation: the meson
field operators are replaced by their expectation values. The sources of the
meson fields are defined by the nucleon densities and currents. The nucleons
move independently in the mean-field potential which originates from the
nucleon-nucleon interaction. The ground state of a nucleus corresponds to
the stationary self-consistent solution of the coupled system of Dirac and
Klein-Gordon equations 
\begin{equation}  \label{statDirac}
\left\{-i\mbox{\boldmath $\alpha$} \cdot\mbox{\boldmath $\nabla$}
+\beta(m+g_\sigma \sigma) +g_\omega \omega^0+g_\rho\tau_3\rho^0_3 +e\frac{
(1-\tau_3)}{2} A^0\right\}\psi_i= \varepsilon_i\psi_i.
\end{equation}
\begin{eqnarray}
\left[ -\Delta +m_{\sigma }^{2}\right] \,\sigma ({\bf r}) &=&-g_{\sigma
}\,\rho _{s}({\bf r})-g_{2}\,\sigma ^{2}({\bf r})-g_{3}\,\sigma ^{3}({\bf r})
\label{messig} \\
\left[ -\Delta +m_{\omega }^{2}\right] \,\omega ^{0}({\bf r}) &=&g_{\omega
}\,\rho _{v}({\bf r})  \label{mesome} \\
\left[ -\Delta +m_{\rho }^{2}\right] \,\rho ^{0}({\bf r}) &=&g_{\rho }\,\rho
_{3}({\bf r})  \label{mesrho} \\
-\Delta \,A^{0}({\bf r}) &=&e\,\rho _{p}({\bf r}),  \label{photon}
\end{eqnarray}
The single-nucleon dynamics is described by the Dirac
equation (\ref{statDirac}). $\sigma$, $\omega$, 
and $\rho$ are the meson fields, and $A$
denotes the electromagnetic potential. $g_\sigma$, $g_\omega$, and $g_\rho$
are the corresponding coupling constants for the mesons to the nucleon. $g_2$
and $g_3$ are the coefficients of the non-linear $\sigma$ terms which
introduce an effective density dependence in the potential. Due to charge
conservation, only the 3rd-component of the isovector rho meson contributes.
The source terms in equations (\ref{messig}) to (\ref{photon}) are sums of
bilinear products of baryon amplitudes with positive energy ({\it no-sea}
approximation).

In addition to the self-consistent mean-field potential, pairing
correlations have to be included in order to describe ground-state
properties of open-shell nuclei. For spherical and deformed nuclei not too
far from the stability line, pairing is often treated phenomenologically in
the simple BCS approximation \cite{Rin.96}. However, the BCS model presents
only a poor approximation for exotic nuclei far from the
valley of $\beta$-stability~\cite{DNW.96}. The structure of
weakly bound nuclei necessitates
a unified and self-consistent treatment of mean-field and pairing
correlations. In particular, the Relativistic Hartree-Bogoliubov (RHB) model
represents a relativistic extension of the Hartree-Fock-Bogoliubov (HFB)
framework. In the RHB model the ground state of a nucleus $\vert \Phi >$ is
represented by the product of independent single-quasiparticle states. These
states are eigenvectors of the generalized single-nucleon Hamiltonian which
contains two average potentials: the self-consistent mean-field $\hat\Gamma$
which encloses all the long range particle-hole ({\it ph}) correlations, and
a pairing field $\hat\Delta$ which sums up the particle-particle ({\it pp})
correlations. In the Hartree approximation for the self-consistent mean
field, the relativistic Hartree-Bogoliubov equations read 
\begin{eqnarray}  \label{rhb}
\left( \matrix{ \hat h_D -m- \lambda & \hat\Delta \cr -\hat\Delta^* & -\hat
h_D + m +\lambda} \right) \left( \matrix{ U_k({\bf r}) \cr V_k({\bf r}) }
\right) = E_k\left( \matrix{ U_k({\bf r}) \cr V_k({\bf r}) } \right).
\end{eqnarray}
where $\hat h_D$ is the single-nucleon Dirac Hamiltonian (\ref{statDirac}),
and $m$ is the nucleon mass. The chemical potential $\lambda$ has to be
determined by the particle number subsidiary condition in order that the
expectation value of the particle number operator in the ground state equals
the number of nucleons. $\hat\Delta $ is the pairing field. The column
vectors denote the quasi-particle spinors and $E_k$ are the quasi-particle
energies. In most applications of the RHB model a phenomenological pairing
interaction has been used: the pairing part of the Gogny force 
\begin{equation}
V^{pp}(1,2)~=~\sum_{i=1,2}e^{-(({\bf r}_{1}-{\bf r}_{2})/{\mu _{i}})^{2}}\,
(W_{i}~+~B_{i}P^{\sigma }-H_{i}P^{\tau }-M_{i}P^{\sigma }P^{\tau }),
\end{equation}
with the set D1S \cite{BGG.84} for the parameters $\mu
_{i}$, $W_{i}$, $B_{i} $, $H_{i}$ and $M_{i}$ $(i=1,2)$.

The RHB equations are solved self-consistently, with potentials determined
in the mean-field approximation from solutions of Klein-Gordon equations for
the meson fields. The Dirac-Hartree-Bogoliubov equations and the equations
for the meson fields are solved by expanding the nucleon
spinors $U_k({\bf r})$ and $V_k({\bf r})$, and the 
meson fields in terms of the eigenfunctions
of a deformed axially symmetric oscillator potential. A detailed description
of the Relativistic Hartree-Bogoliubov model for deformed nuclei can be
found in Ref. \cite{LVR.99}. 
%=========================================================================
%  Section 3
%

\section{Shape coexistence in neutron-deficient Hg and Pb isotopes}

%=========================================================================
The light isotopes of Hg and Pb exhibit a variety of
coexisting shapes~\cite{Wood.92,Jul.01}. 
A systematic analysis of shape coexistence effects at $%
I^{\pi} = 0^+$ in these nuclei has been performed by 
Nazarewicz~\cite{Naz.93}, using the shell 
correction approach with the axial, reflection asymmetric
Woods-Saxon model. A common feature is the competition of low-lying prolate
and oblate shapes. The ground states of Hg isotopes are weakly oblate
deformed (two-proton hole states). In $^{188}$Hg the oblate ground-state
band is crossed by the intruding deformed band which corresponds to a
prolate minimum ($4p-6h$ proton excitations into the $h_{9/2}$ and $f_{7/2}$
orbits). The excitation energy of the prolate band is lowered with
decreasing neutron number, and reaches a minimum in $^{182}$Hg. For lighter
Hg isotopes the excitation energy of the prolate minimum increases, and the
oblate ground state evolves toward a spherical shape. Experimental data on
energy spectra and charge radii show that the ground states of Pb isotopes
are spherical, but both oblate ($2p-2h$ proton excitations)
and prolate ($4p-4h$ proton 
excitations) low-lying minima are observed for $N < 110$. A
beautiful example of oblate and prolate minima at almost identical
excitation energies is found in $^{186}$Pb~\cite{ANA.00}. A recent review of
experimental data on intruder states in neutron deficient Hg, Pb and Po
nuclei can be found in Ref.~\cite{Jul.01}.

In Ref.~\cite{Ben.01} Bender {\it et al.} have used non-relativistic and
relativistic self-consistent mean field models to analyze recent
experimental data which seem to indicate a strong reduction of the $Z=82$
shell gap in neutron-deficient Pb isotopes. The systematics of differences
between two-proton separation energies in adjacent even-even nuclei with the
same neutron number (two-proton shell gaps)~\cite{Nov.02}, $Q_{\alpha}$
values and $\alpha$ reduced widths~\cite{Toth.99} suggest a lessened
magicity of the $Z=82$ shell when neutrons are midway
between $N=82$ and $N=126$. 
However, in Ref.~\cite{Ben.01} it has been shown that the
systematics of two-proton shell gaps can be described quantitatively in
terms of deformed ground states of Hg and Po isotopes. The
experimental 
$Q_{\alpha}$ values also reflect 
deformation effects, and the systematics of 
$\alpha$-decay hindrance factors is consistent with the
stability of the $Z=82 $ shell gap.

The structure of coexisting minima in neutron-deficient Hg and Pb nuclei has
also been analyzed in the relativistic mean-field
framework. In Refs.~\cite{Yos.94,Pat.94} 
the RMF model, with the NL1~\cite{RRM.86} effective
interaction, was used to calculate prolate, oblate and spherical solutions
for neutron-deficient Pt, Hg and Pb isotopes. Pairing was treated in the BCS
approximation with constant pairing gaps. Although the NL1 effective force
strongly over binds these nuclei, i.e. the calculated
binding energies are $10-12$ MeV larger than the empirical
ones, nevertheless detailed predictions
were made for ground state deformations and relative positions of coexisting
minima, including the excitation energies of superdeformed states. Most of
these results, as pointed out in Ref.~\cite{Hey.96}, contradict well
established experimental data (wrong sign of the ground state deformations
for Hg isotopes, calculated deformed ground states for some Pb nuclei,
superdeformed ground states, etc.). The calculations were
repeated in Ref.~\cite{Tak.96} with the NL-SH effective 
interaction~\cite{SNR.93}, and the
values of the pairing gaps were varied by as much as 50\%, but the results
remained essentially the same. Eventually, deformed RMF+BCS calculations
with the NL1 effective interaction reproduced the experimental ground state
oblate deformations of $^{180-188}$Hg~\cite{Yos.97}. The differences of the
intrinsic energies of the lowest prolate and oblate states were also in
agreement with experimental data. The gap parameters were not kept
constant in each nucleus, rather they were determined in a 
self-consistent way from a monopole force with constant strength
parameter $G$. This leads to deformation dependent pairing gaps.
In particular, for the oblate ground states the proton
pairing gaps vanish, while the neutron pairing gap is a factor two larger
than the average value $12/\sqrt{A}$. Since the calculated ground-state
properties are very sensitive to both pairing and deformation effects, it
was argued~\cite{Yos.97} that a unified and self-consistent framework, i.e.
the relativistic Hartree-Bogoliubov model, might be more appropriate for the
study of neutron-deficient nuclei in this mass region. It is not only
the treatment of pairing which  makes it difficult to assess the
conclusions of Ref.~\cite{Yos.97},  but also the large deviations of the
calculated binding energies from the experimental values. The relative
positions of prolate and oblate minima differ by only $\approx 0.5$ MeV in
most cases, while the calculated binding energies are more than 10 MeV too
large. This is, of course, caused by the well known fact that the parameter
set NL1 fails to reproduce nuclear binding energies far from stability.

In this work we calculate ground state properties of neutron-deficient Hg
and Pb nuclei in the relativistic Hartree-Bogoliubov framework. In most
applications of the RHB model we have used the NL3
effective interaction~\cite{LKR.97} 
in the particle-hole channel, and pairing correlations were
described by the pairing part of the finite range Gogny
interaction D1S~\cite{BGG.84}. This force has been very 
carefully adjusted to pairing properties
of finite nuclei all over the periodic table. In particular, the basic
advantage of the Gogny force is the finite range, which automatically
guarantees a proper cut-off in momentum space. Properties calculated with
NL3 indicate that this is probably the best effective relativistic
interaction so far, both for nuclei at and away from the
line of $\beta $-stability. 
A recent systematic theoretical study of ground-state properties
of more than 1300 even-even isotopes has shown very good agreement with
experimental data \cite{LRR.98}. However, in the analysis of the $Z=82$
shell closure in neutron-deficient Pb isotopes~\cite{Ben.01}, it was noted
that RMF+BCS calculations with the NL3 interaction predict deformed ground
states for several Pb nuclei.

In Fig.~\ref{figA} we display the calculated binding energy curves of even-A 
$^{184-198}$Pb isotopes as functions of the quadrupole deformation. The
curves correspond to axially deformed RHB model solutions with constrained
quadrupole deformation. The effective interaction is NL3 + Gogny D1S. $^{184}
$Pb and $^{186}$Pb have spherical ground states, and we find low-lying
oblate and prolate minima, in qualitative agreement with
experimental data~\cite{Jul.01}. 
With increasing neutron number, however, the oblate minimum
is lowered in energy and the nuclei $^{188-194}$Pb have oblate ground
states. This result, of course, contradicts experimental
data. Only from $A=196$ the 
calculated Pb ground states become again spherical, but
both for $^{196}$Pb and $^{198}$Pb the potential curves
display wide and flat minima
on the oblate side.

The RHB NL3+D1S binding energy curves of even-A $^{176-190}$Hg nuclei are
shown in Fig.~\ref{figB}. Model calculations predict a spherical ground
state for $^{176}$Hg and an almost spherical, but slightly oblate ground
state for $^{178}$Hg. With increasing neutron number the spherical state
evolves into an oblate minimum, and a pronounced minimum develops on the
prolate side. The positions of the minima, $\beta_2 \approx -0.15$ on the
oblate side, and $\beta_2 \approx 0.3$ for the prolate minimum, agree well
with the values calculated by Nazarewicz~\cite{Naz.93} using the
Nilsson-Strutinsky approach. The problem is, however, the relative
excitation energies. For the NL3+D1S effective interaction the prolate
minimum is the ground state for $^{180,182,184,186}$Hg. The calculated
ground states are oblate for $^{188,190}$Hg. These results are at variance
with experimental data: all Hg isotopes with $A \geq 178$ have oblate ground
states, the prolate minimum is estimated at $300 - 800$ keV above the ground
state~\cite{Jul.01}. If the NL1 effective interaction is used in the $ph$
channel, we find that the oblate minima are always lower, and this is
similar to the result obtained with the RMF+BCS calculation
of Ref.~\cite{Yos.97}. 
The calculated binding energies, however, are on the average more
than 10 MeV larger than the experimental ones.

Is it then possible to construct a relativistic effective interaction which
will consistently describe the ground state properties of neutron-rich
nuclei in this mass region? The motivation, of course, is not just to
reproduce the experimental data. The real question is whether effective
mean-field interactions can be accurately extrapolated from stable nuclei to
isotopes with extreme isospin values and to the drip lines. In the
particular example considered in the present study, the quantitative
description of coexisting spherical, oblate and prolate minima is, of
course, beyond the mean-field approach. A detailed analysis of coexisting
shapes necessitates the use of models which can account for configuration
mixing effects. However, a consistent description of ground states should be
possible in the relativistic mean-field framework, even if their properties
are affected by correlations not explicitly included in the model. In a
series of recent papers (see, for example, Refs.~\cite{FS.00,FS.01}),
Furnstahl and Serot have argued that relativistic mean-field models should
be considered as approximate implementations of the Kohn-Sham density
functional theory, with local scalar and vector fields appearing in the role
of relativistic Kohn-Sham potentials. The mean-field models approximate the
exact energy functional of the ground state density of a many-nucleon
system, which includes all higher-order correlations, using powers of
auxiliary meson fields or nucleon densities.

In the recent analysis of the phenomenon of shape coexistence within the
non-relativistic self-consistent Hartree-Fock method and the nuclear shell
model~\cite{PG.99}, it was emphasized that the sizes of spherical magic and
semi-magic gaps in the single-nucleon spectrum determine the relative
positions of many-particle many-hole intruder configurations with respect to
the ground state. The spherical gaps are the main factor that determines the
relative excitation energies of coexisting minima based on different
intruder configurations. The relative distance between the individual shells
also determines the effective strength of the quadrupole 
field~\cite{Naz.94}. 
The spherical proton $Z=82$ shell closure is illustrated in
Fig.~\ref{figC}, where we display the 
last four occupied and the first three unoccupied
proton single-particle states in $^{208}$Pb. The experimental levels in the
first column are from Ref. ~\cite{Lev}. The levels in the second, third and
fourth columns are calculated in the RHB model with the effective
interactions NL1, NL-SC, and NL3, respectively. Comparing
the magic gaps $E(h_{9/2}) - E(s_{1/2})$, 
one notices that the value calculated with the NL3
effective interaction is much smaller than the empirical gap. This explains
why several neutron-deficient Pb isotopes, calculated with the NL3
interaction, have deformed ground states. The magic gap in the proton
spectrum calculated with NL1, on the other hand, is somewhat larger than the
empirical value. The spectrum of proton occupied states is, however, more
bound and the total binding energies calculated with NL1 do not compare well
with the experimental masses of lighter Pb isotopes. We have tried,
therefore, to construct a new effective interaction which, on one hand,
preserves the good properties of NL3 (binding energies, charge radii,
isotopic shifts), but at the same time takes into account the gap between
the last occupied and the first unoccupied major spherical shells. The
result is shown in the third column NL-SC of Fig.~\ref{figC} (SC for shape
coexistence). There are several comments which should be made at this point.

First, the NL-SC interaction has been constructed by starting from the
standard NL1 parameterization, and changing the parameters towards NL3.
While the masses and coupling constants of the $\sigma$ and $\omega$ mesons
are those of NL1, the parameters of the non-linear $\sigma$-terms and,
especially important for the binding energies of isotopes with small N/Z,
the coupling constant of the $\rho$ meson, are very close to the NL3
parameter set. It should also be emphasized that, although in the
construction of the NL-SC interaction we paid particular attention to the
Pt-Hg-Pb region, the parameters were, as usual, checked to reproduce a set
of ground state data of about ten spherical nuclei from different mass
regions. For nuclei with $A > 100$ the calculated $\chi^2$ per datum is
comparable to the one obtained with the standard NL3 interaction. The ground
state properties of light nuclei calculated with NL-SC, on the other hand,
are not as good as with NL3 and therefore we do not regard NL-SC as a new
general effective relativistic mean field interaction. NL-SC has been
specifically tailored to illustrate the problem considered in the present
analysis.

Second, the construction of the NL-SC interaction goes beyond the standard
mean-field approach, in the sense that the parameters of the mean-field
functional depend not only on the ground-state density, i.e. on the occupied
states, but implicitly also on the relative position of the unoccupied
spherical shells. In order to be consistent with the mean field approach,
the calculated spherical gap should be larger than the empirical value. It
is well known that the coupling of single particle states to collective
vibrations increases the effective mass in the vicinity of the Fermi level,
and this effect decreases the gap between occupied and unoccupied shells. We
have recently shown how to include the particle-vibration coupling in the
self-consistent relativistic mean-field framework~\cite{VNR.02}, but the
present analysis does not consider this effect. In Ref.~\cite{PG.99} it was
emphasized that the proper treatment of pairing and zero-point correlations,
vibrational and rotational, is crucial for detailed predictions of shape
coexistence effects. And while we treat pairing and mean-field correlations
in the unified Hartree-Bogoliubov framework, no attempt is made to
explicitly include zero-point energy corrections. Since in the analysis of
Ref.~\cite{PG.99} it was shown that the sum of rotational and vibrational
zero-point energy corrections is approximately constant as function of
deformation for a given nucleus, we assume that these corrections can be
absorbed in the parameters of an effective interaction adjusted to reproduce
binding energies, at least in a limited mass region. In any case, since the
effective interaction NL-SC is adjusted to reproduce the spherical proton
shell gap at $Z=82$, one cannot really expect that it accurately describes
the ground state shapes of nuclei in other (Z,N) regions characterized by
shape coexistence.

In Figs.~\ref{figD} and \ref{figE} we display the binding energies per
nucleon of even-A Hg and Pb isotopes, respectively, calculated in the RHB
model with the mean-field effective interactions NL1, NL3, and NL-SC, and
with the Gogny D1S interaction in the pairing channel. The theoretical
binding energies are compared with the empirical values
from Ref.~\cite{AW.95}. 
Both the NL3 and the NL-SC interactions reproduce in detail the
mass dependence of the binding energies of Pb isotopes, while the NL1
interaction strongly over binds the nuclei below $^{208}$Pb. The reason is
the much larger value of the $\rho$ meson coupling constant. Essentially
identical binding energies are calculated with all three forces for nuclei
above $^{208}$Pb. A similar effect is observed for the Hg
isotopes in Fig.~\ref{figD}. 
While NL3 and NL-SC reproduce much better the empirical data for
the neutron-deficient Hg isotopes, comparable results are obtained with all
three interactions for $A\geq 100$.

A remarkable success of relativistic mean field models was the realization
that, because of the intrinsic isospin dependence of the effective
single-nucleon spin-orbit potential, they naturally reproduce the anomalous
charge isotope shifts~\cite{SLR.93}. The well known example of the anomalous
kink in the isotope shifts of Pb isotopes is shown in Fig.~\ref{figF}. The
results of RHB calculations with the NL1, NL3, and NL-SC effective
interactions, and with the Gogny D1S interaction in the pairing channel, are
compared with experimental data from Ref.~\cite{Rad}. While all three forces
reproduce the general trend of isotope shifts and the kink at $^{208}$Pb,
this effect is more pronounced for NL1 and NL3. The experimental data are
best reproduced by the NL-SC interaction. The RHB theoretical values for the
charge isotope shifts of even-A Hg isotopes are compared with the
experimental data~\cite{Rad} in Fig.~\ref{figG}. The large discrepancy
between data and the values calculated with NL3 for $A < 188$, simply
reflects the fact that this interaction predicts the wrong sign for the
ground state deformations of neutron-deficient Hg nuclei. The interactions
NL1 and NL-SC reproduce the experimental data, although the agreement is not
as good as in the case of Pb isotopes.

In Fig.~\ref{figH} we plot the differences between the binding energies of
prolate and oblate minima in even-A $^{180-188}$Hg isotopes. The values
calculated with the NL1, NL3, and NL-SC effective interactions are shown in
comparison with the empirical data~\cite{Diff}. As it was already shown in
Fig.~\ref{figB}, prolate ground states are calculated with the NL3
interaction. The NL1 and NL-SC interactions, on the other hand, reproduce
the relative positions of the prolate and oblate minima. One has to keep in
mind, however, that the total binding energies calculated with NL1 are, on
the average, more than 10 MeV too large. The RHB model with the NL-SC
interaction in the $ph$ channel and with the Gogny D1S
interaction in the $pp $ channel, 
provides a consistent description for the ground state
properties of neutron-deficient nuclei in the Pt-Hg-Pb region (binding
energies, radii, ground state deformations). We should emphasize that NL-SC
was not adjusted to reproduce the relative excitation energies of the
prolate minima in Hg nuclei. However, the fact that the calculated relative
positions of prolate and oblate minima agree with the empirical values,
might indicate that configuration mixing effects are not very important in
these nuclei.

%=========================================================================
%  Section 4
%

\section{Conclusions}

%=========================================================================
This work presents an analysis of ground-state properties of
neutron-deficient Hg and Pb isotopes in the framework of the Relativistic
Hartree-Bogoliubov (RHB) model. In the last couple of years this model has
been very successfully applied in the description of nuclear structure
phenomena in medium-heavy and heavy exotic nuclei far from
the valley of $\beta$- stability and of the 
physics of the drip lines. It is still a very
much open problem, however, how far from stability can one apply
relativistic effective interactions which have been adjusted to global
properties of a small number of spherical closed shell nuclei. Can these
interactions accurately describe experimentally known properties and predict
new structure phenomena in exotic nuclei far from stability?

One of the testing grounds of nuclear structure models is their ability to
describe shape coexistence phenomena in soft nuclei. In the present study,
in particular, we have analyzed the results of relativistic mean-field
calculations in the region of neutron deficient Pt-Hg-Pb nuclei. Standard
relativistic interactions, adjusted to ground state properties of stable
nuclei, do not reproduce the empirical ground state shapes of neutron
deficient nuclei in this mass region. We have shown, however, that effective
interactions can be constructed, which consistently describe binding
energies, charge radii, ground state quadrupole deformations and, at least
qualitatively, the relative positions of coexisting minima in Hg and Pb
isotopic chains. In adjusting the parameters of such an interaction, in
addition to the usual experimental data on ground states of stable nuclei,
also the sizes of spherical gaps in the single-nucleon spectra have to be
taken into account. The spherical magic and semi-magic gaps determine the
relative excitation energies of coexisting minima based on different
intruder configurations.

Although a quantitative analysis of shape coexistence phenomena goes beyond
the mean-field approach, a consistent description of ground states is
possible in the relativistic mean-field framework. We have shown that the
RHB model with the NL-SC interaction in the $ph$ channel, 
adjusted to the $Z=82$ proton shell closure in $^{208}$Pb, and with the Gogny
D1S interaction in the $pp$ channel, reproduces in detail the empirical
ground state properties of neutron-deficient Hg and Pb nuclei. 

The results of the present analysis suggest that, when constructing
effective mean-field interactions to be used in regions of exotic nuclei far
from stability, not only bulk properties of spherical nuclei, but also magic
and semi-magic gaps in the single nucleon spectra must be taken into
account. \bigskip \bigskip

\leftline{\bf ACKNOWLEDGMENTS}

This work has been supported in part by the Bundesministerium f\"{u}r
Bildung und Forschung under project 06 TM 979, and by the
Gesellschaft f\"{u}r Schwerionenforschung (GSI) 
Darmstadt. T.N. and D.V. would like to
acknowledge the support from the Alexander von Humboldt - Stiftung. 

\newpage 
\begin{figure}[tbp]
\caption{Binding energy curves of even-A Pb isotopes as functions of the
quadrupole deformation. The curves correspond to RHB model solutions with
constrained quadrupole deformation. The effective interaction is NL3 + Gogny
D1S.}
\label{figA}
\end{figure}

\begin{figure}[tbp]
\caption{Same as in Fig.~\protect\ref{figA}, but for neutron-deficient Hg
isotopes.}
\label{figB}
\end{figure}

\begin{figure}[tbp]
\caption{Proton single-particle states in $^{208}$Pb. The experimental
levels in the first column are from Ref. ~\protect\cite{Lev}. The levels in
the second, third and fourth columns are calculated in the RHB model with
the effective interactions NL1, NL-SC, and NL3, respectively.}
\label{figC}
\end{figure}

\begin{figure}[tbp]
\caption{Binding energies of even-A Hg isotopes calculated in the RHB model
with the mean-field effective interactions NL1, NL3, and NL-SC, and with the
Gogny D1S interaction in the pairing channel. The theoretical binding
energies are compared with the empirical values from 
Ref.~\protect\cite{AW.95}.}
\label{figD}
\end{figure}

\begin{figure}[tbp]
\caption{Same as in Fig.~\protect\ref{figD}, but for Pb isotopes.}
\label{figE}
\end{figure}

\begin{figure}[tbp]
\caption{Charge isotope shifts in even-A Pb isotopes. The results of RHB
calculations with the NL1, NL3, and NL-SC effective interactions, and with
the Gogny D1S interaction in the pairing channel, are compared with
experimental data from Ref. ~\protect\cite{Rad}.}
\label{figF}
\end{figure}

\begin{figure}[tbp]
\caption{The RHB theoretical values for the charge isotope shifts in even-A
Hg isotopes, compared with experimental data from Ref. ~\protect\cite{Rad}.}
\label{figG}
\end{figure}

\begin{figure}[tbp]
\caption{The difference between binding energies of prolate and oblate
states in even-A Hg isotopes. The results of RHB calculations with the NL1,
NL3, and NL-SC effective interactions, and with the Gogny D1S interaction in
the pairing channel, are compared with the empirical data
from Ref. ~\protect\cite{Diff}.}
\label{figH}
\end{figure}

%%%%%%%%%%%%%%%%%%%%%%%%%%%%%%%%%%%%%%%%%%%%%%%%%%%%%%%%%%%%%

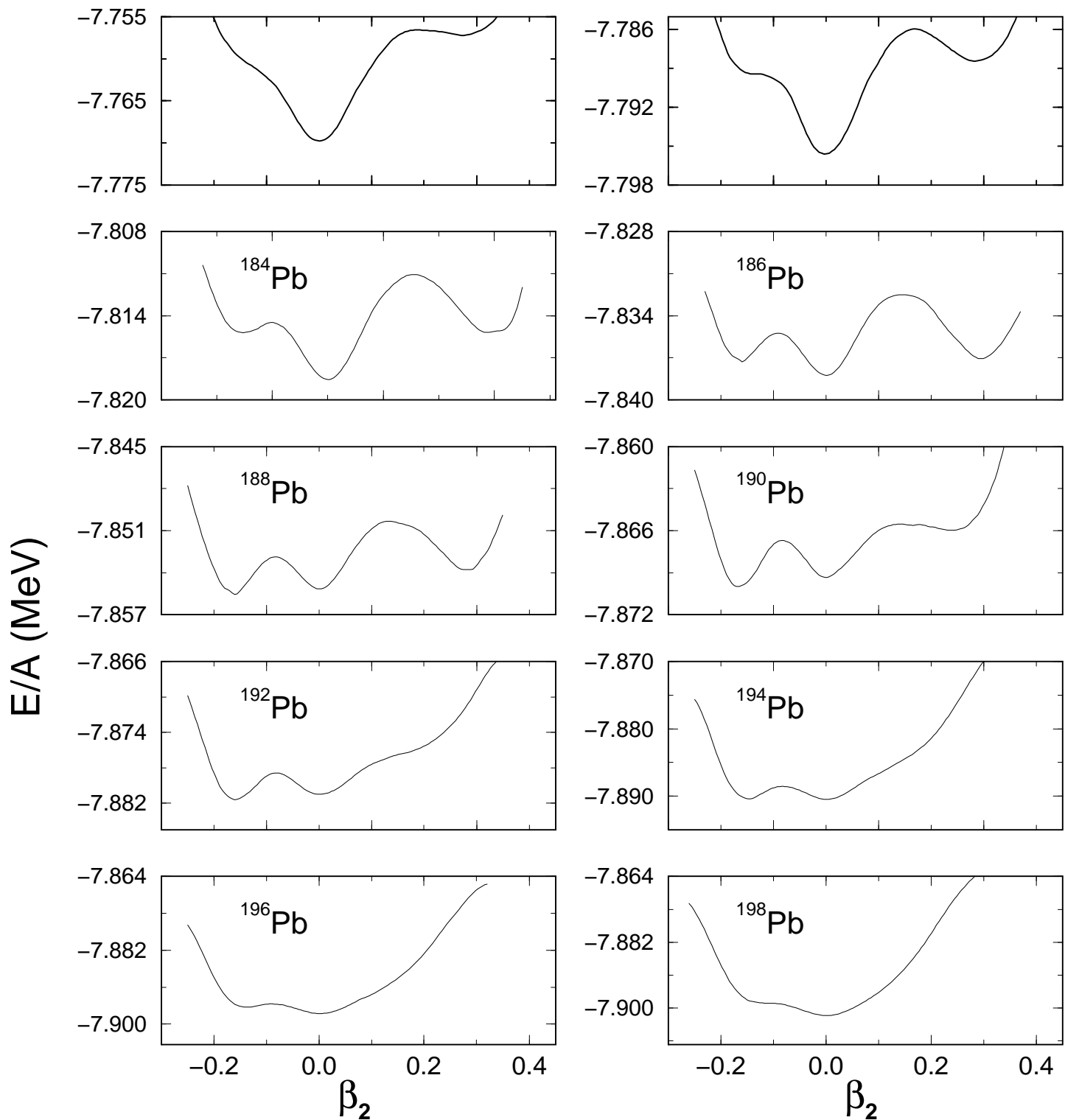
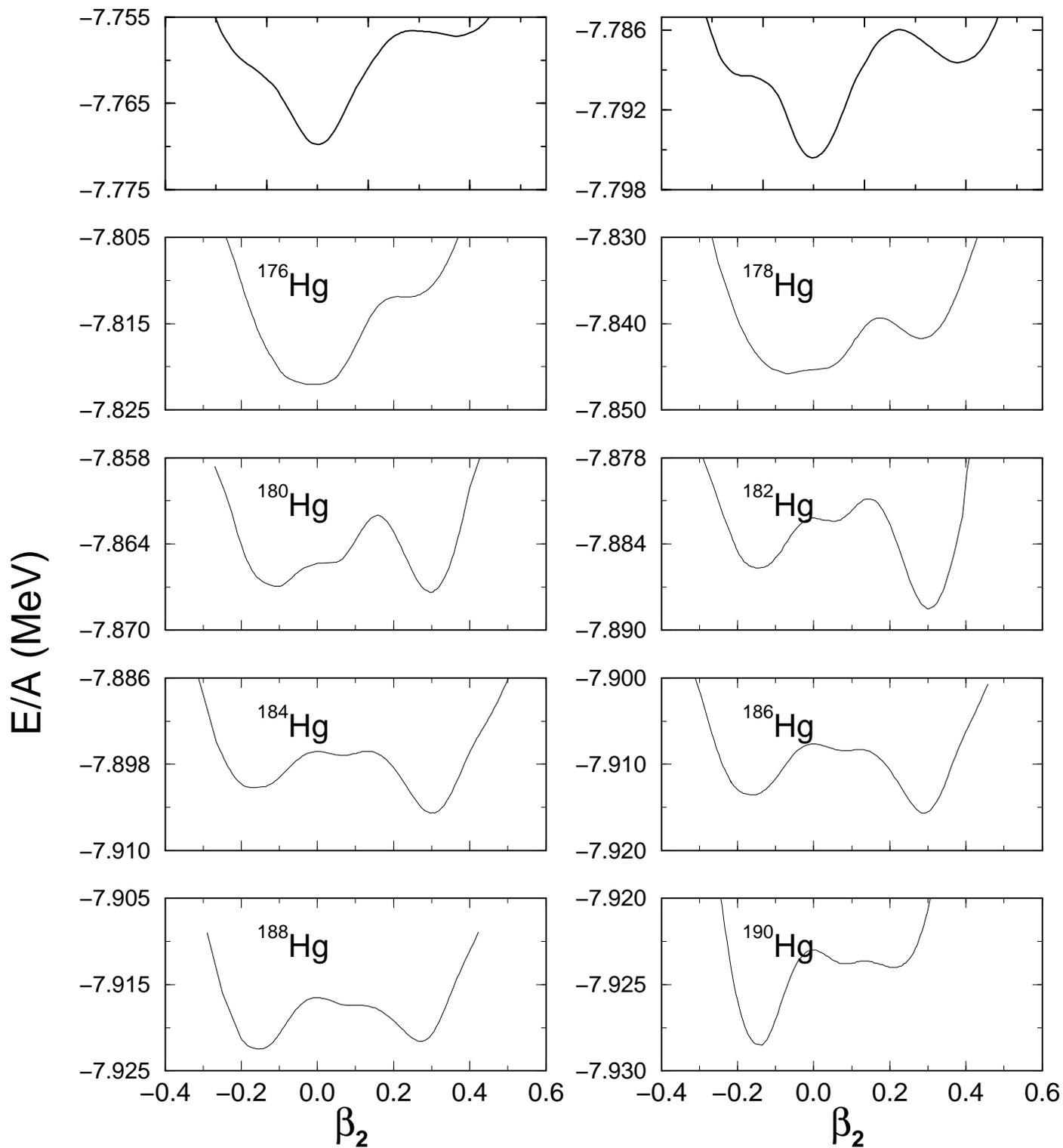
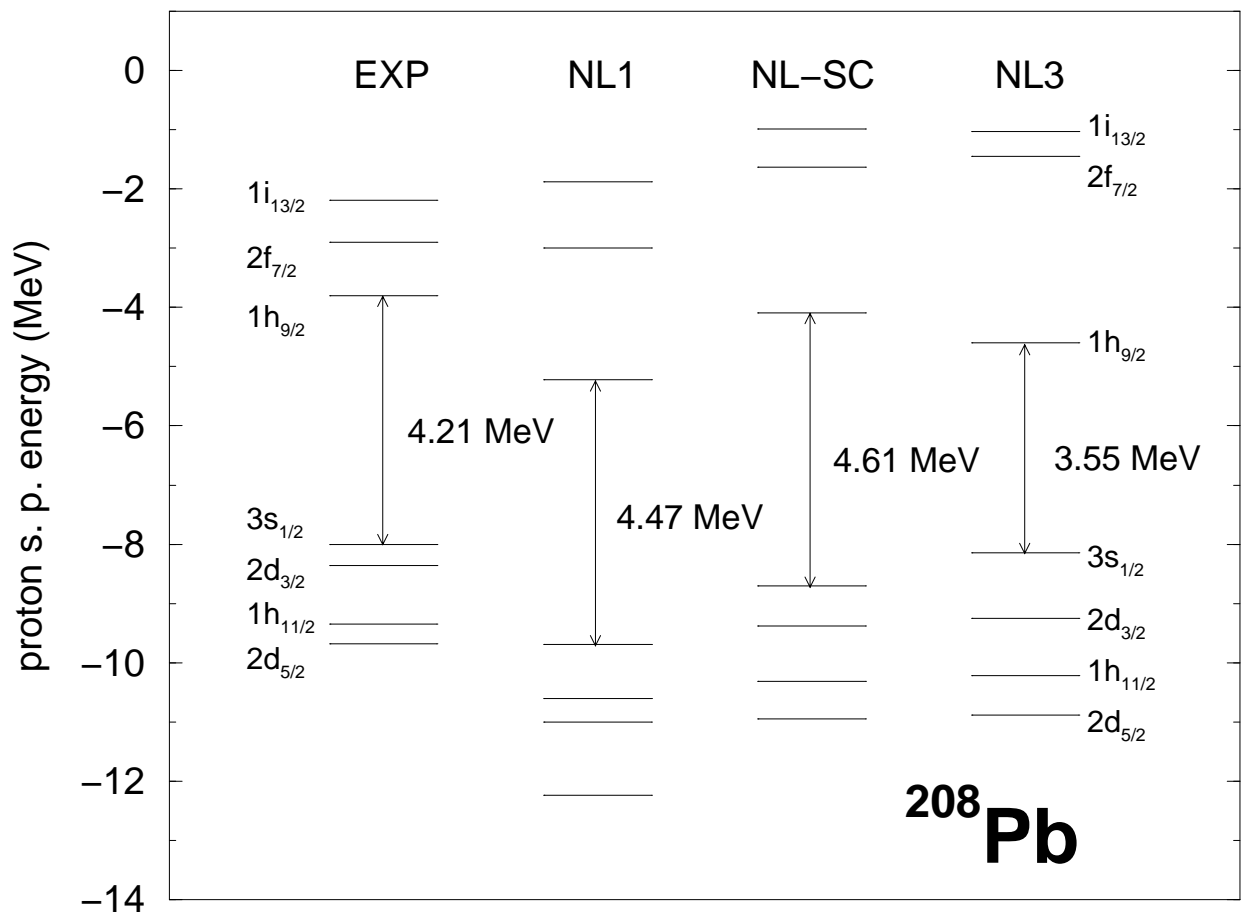
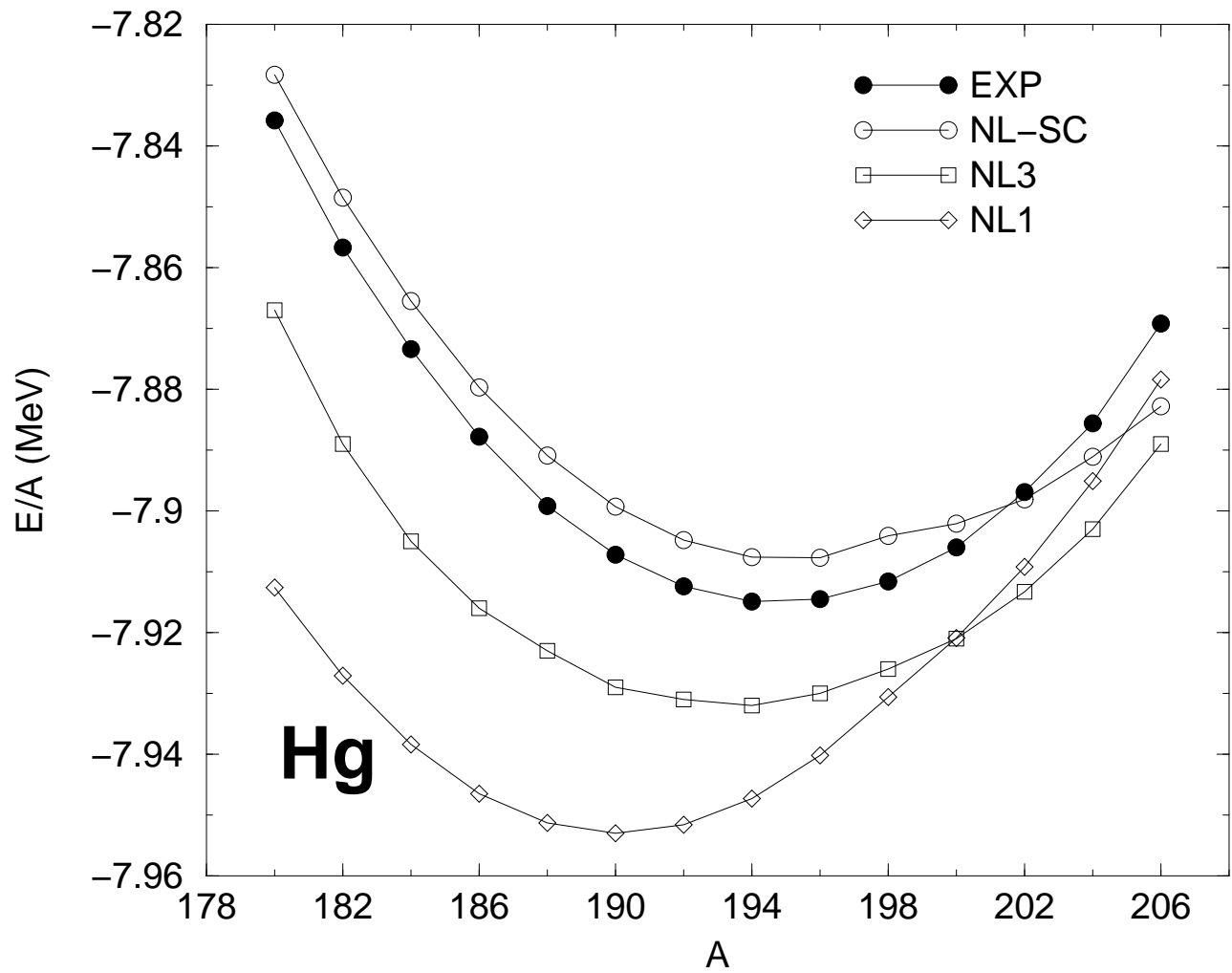
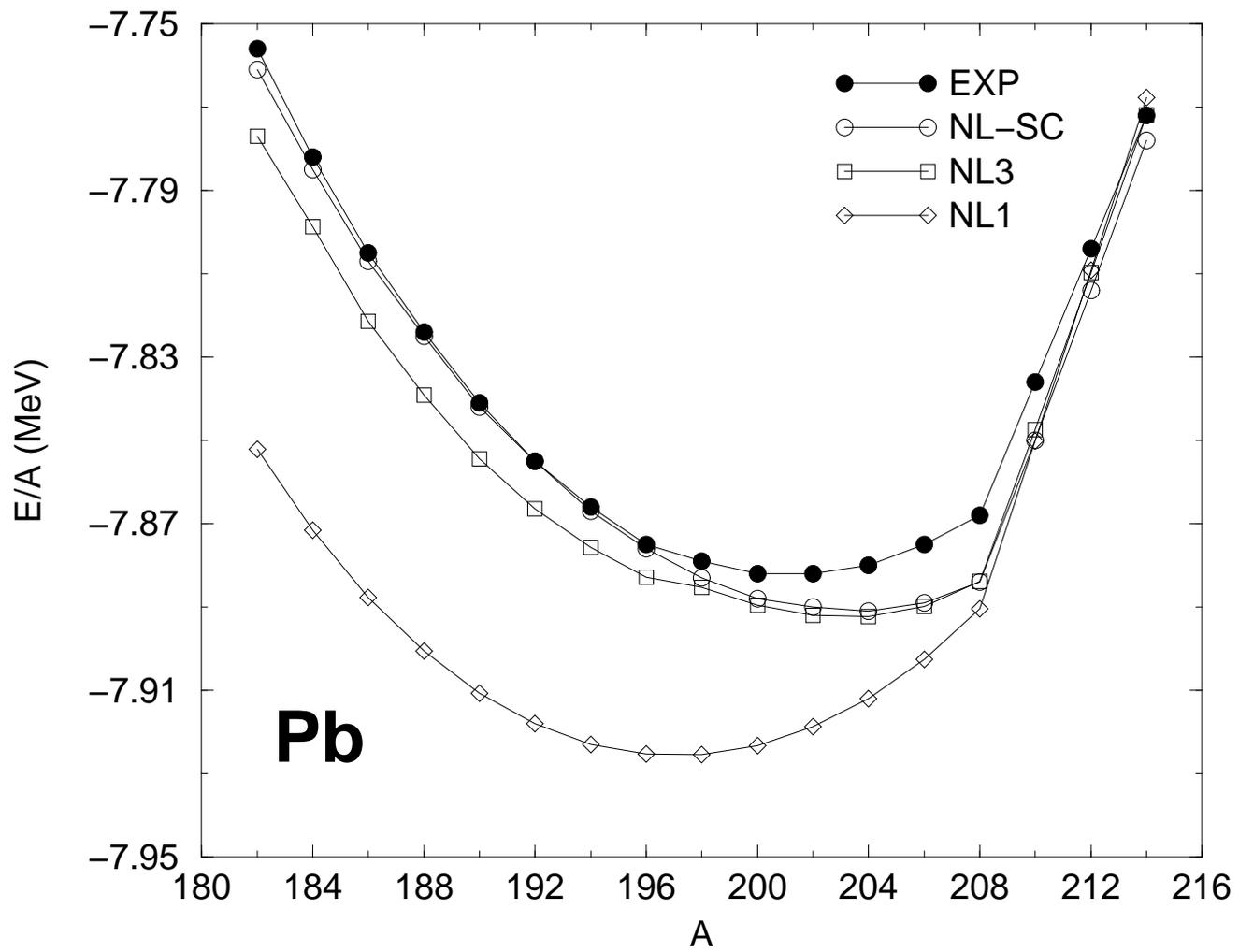
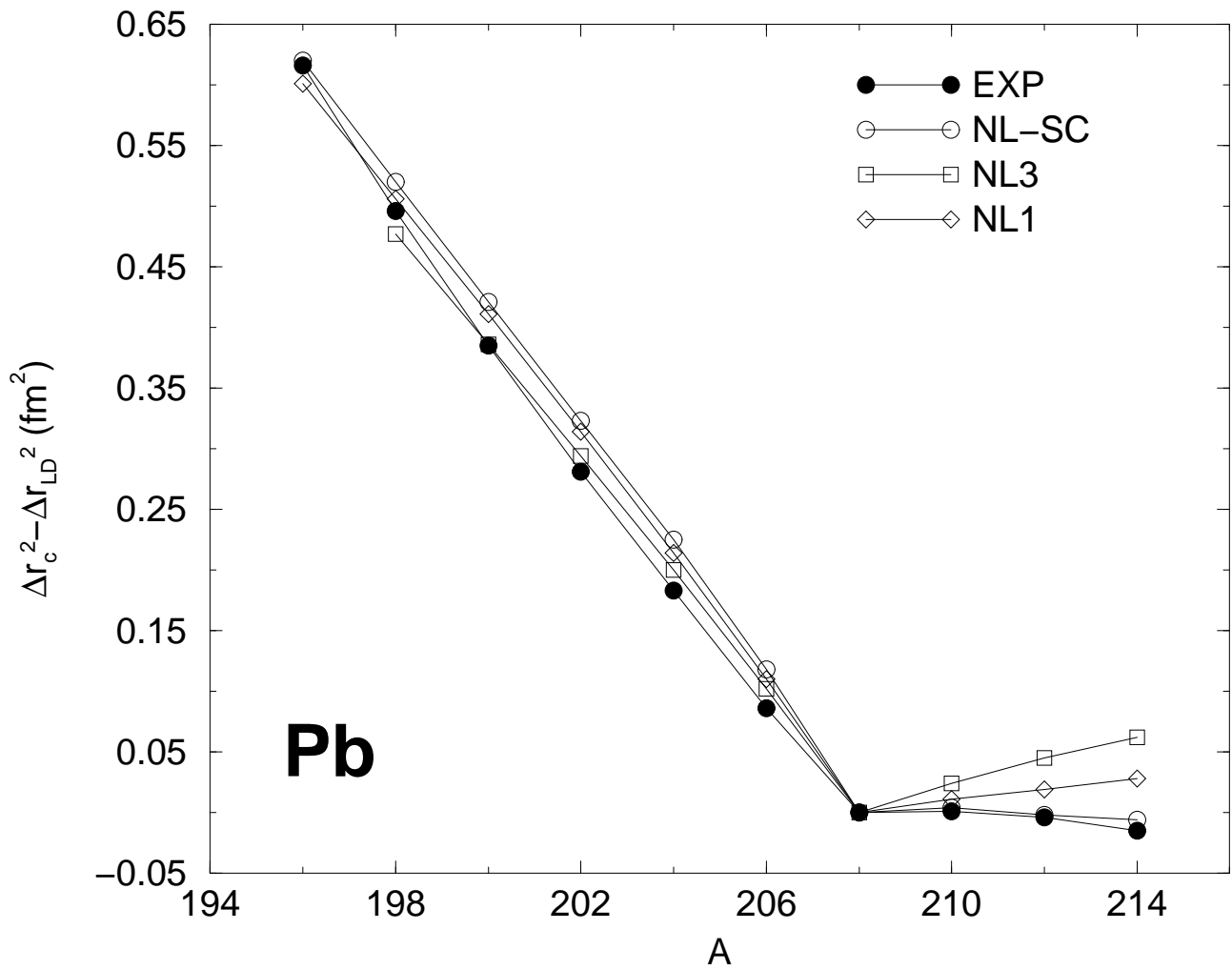
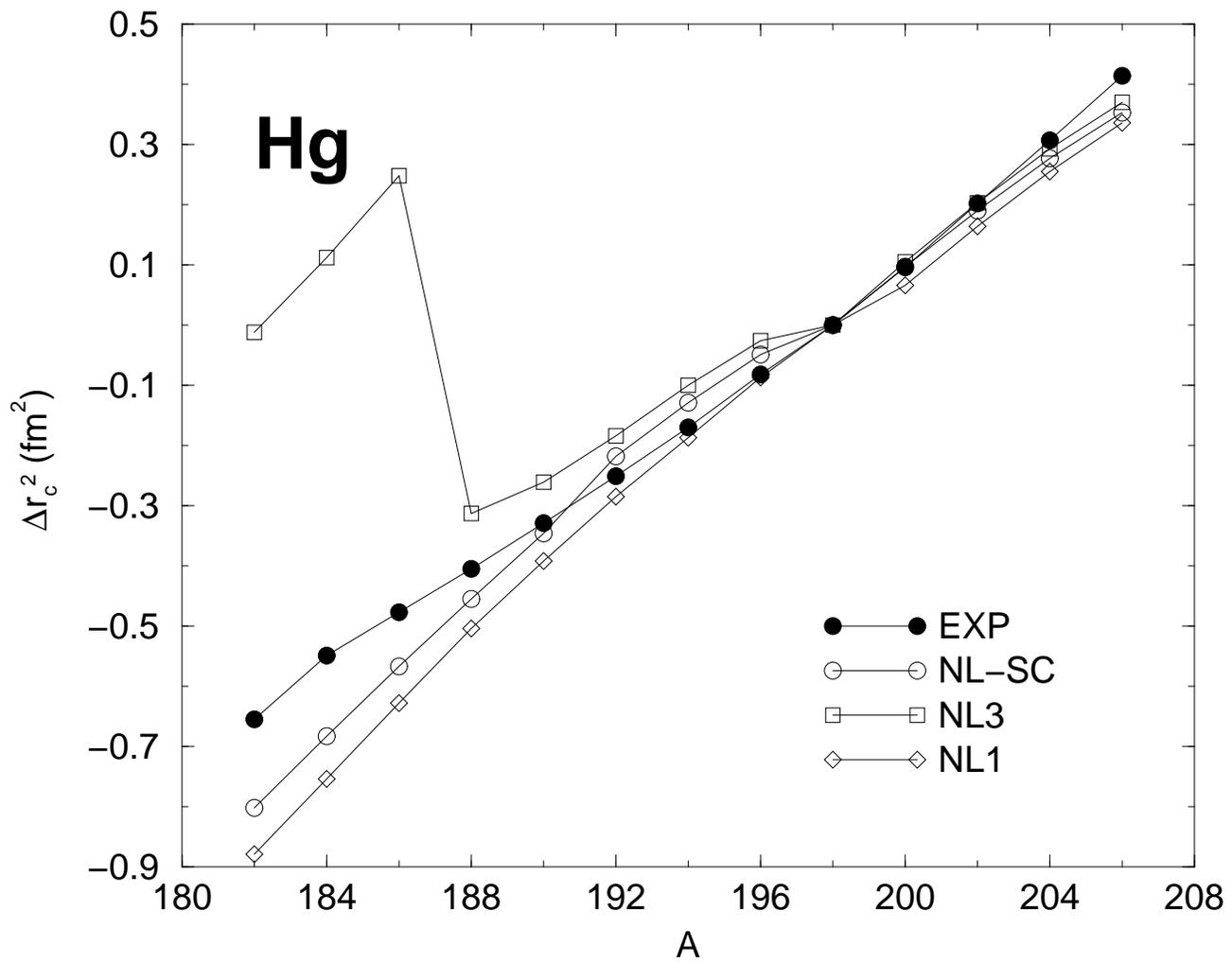
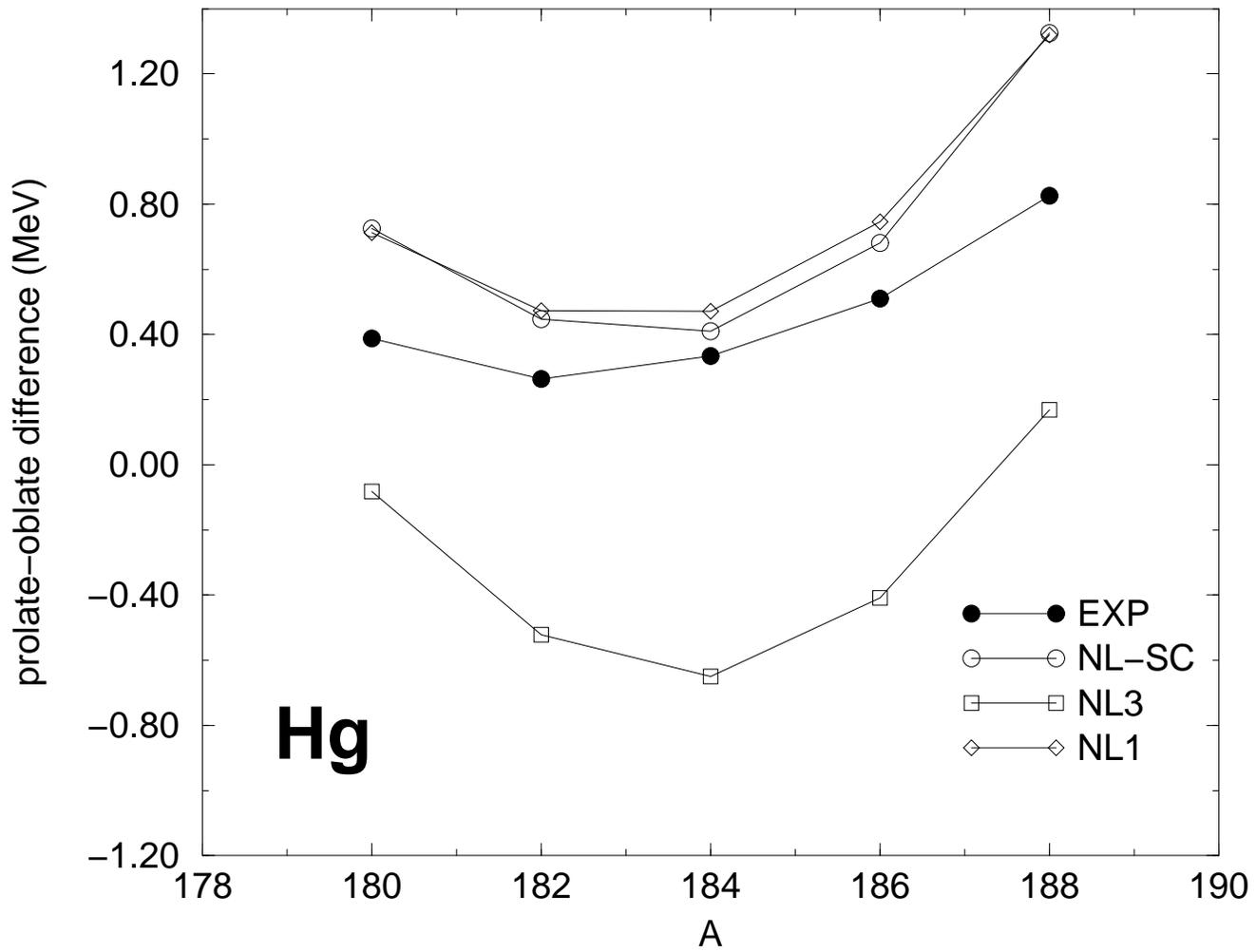
\end{document}